\documentclass[fleqn,usenatbib]{mnras}

\usepackage[T1]{fontenc}
\usepackage{pdflscape}

\DeclareRobustCommand{\VAN}[3]{#2}
\let\VANthebibliography\thebibliography
\def\thebibliography{\DeclareRobustCommand{\VAN}[3]{##3}\VANthebibliography}

\usepackage{graphicx}	
\usepackage{amsmath}	
\usepackage{amssymb}	

\title[Local drivers of galaxy evolution with MaNGA]
      {Probing the major driver of stellar population properties over sub-galaxy scales with SDSS
      MaNGA IFU spectroscopy}

\author[Ferreras et al.]
{I. Ferreras$^{1,2,3}$\thanks{E-mail: i.ferreras@ucl.ac.uk}, 
M. Trevisan$^{4}$, O. Lahav$^1$, R.~R. de Carvalho$^5$, J. Silk$^{6,7,8}$\\
$^1$ Department of Physics and Astronomy, University College London, London WC1E 6BT, UK\\
$^2$ Instituto de Astrof{\'i}sica de Canarias, Calle V{\'i}a L{\'a}ctea s/n,
E38205, La Laguna, Tenerife, Spain\\
$^3$ Departamento de Astrof{\'i}sica, Universidad de La Laguna, E38206 La Laguna, Tenerife, Spain\\
$^4$ Departamento de Astronomia, Universidade Federal do Rio Grande do Sul, 91501-970 Porto Alegre-RS, Brazil\\
$^5$ NAT - Universidade Cidade de S\~ao Paulo, 01506-000, SP, Brazil\\
$^6$ Institut d'Astrophysique de Paris, 98 bis Boulevard Arago, F-75014 Paris, France\\
$^7$ BIPAC, Department of Physics, University of Oxford, Keble Road, Oxford OX1 3RH, UK\\
$^8$ Department of Physics and Astronomy, The Johns Hopkins University, Baltimore, MD 21218, USA\\
}

\date{Submitted for publication in MNRAS, December 4th, 2024}

\pubyear{2024}

\begin{document}
\label{firstpage}
\pagerange{\pageref{firstpage}--\pageref{lastpage}}
\maketitle

\begin{abstract}
Thanks to Integral Field Unit survey data it is possible to explore in detail the link between the formation of the stellar content in galaxies and the drivers of evolution. Traditionally, scaling relations have connected galaxy-wide parameters such as stellar mass (M$_s$), morphology or average velocity dispersion ($\sigma$) to the star formation histories (SFHs). We study a high quality sample of SDSS-MaNGA spectra to test the possibility that sub-galaxy ($\sim$2\,kpc) scales are dominant, instead of galaxy-wide parameters. We find a strong correlation between {\sl local} velocity dispersion and key line strengths that depend on the SFHs, allowing us to make the ansatz that this indicator -- that maps the local gravitational potential -- is the major driver of star formation in galaxies, whereas larger scales play a role of a secondary nature. Galactocentric distance has a weaker correlation, suggesting that the observed radial gradients effectively reflect local variations of velocity dispersion.  In our quest for a cause, instead of a correlation, we contrast $\sigma$ with local stellar mass, that appears less correlated with population properties.  We conclude that the inherently higher uncertainty in M$_s$ may explain its lower correlation with respect to $\sigma$, but the extra uncertainty needed for $\sigma$ to have similar correlations as M$_s$ is rather high. Therefore we posit local velocity dispersion as the major driver of evolution, a result that should be reproduced by hydrodynamical models at the proper resolution.
\end{abstract}

\begin{keywords}
  galaxies: evolution -- galaxies: stellar content -- galaxies: statistics -- galaxies: fundamental parameters -- techniques: spectroscopic -- methods: data analysis
\end{keywords}


\section{Introduction}
\label{Sec:Intro}

Over the past two decades, large scale spectroscopic surveys have
revolutionised our understanding of extragalactic astrophysics.
Before these large datasets became available, most studies were
based on relatively small numbers of well-chosen targets. Survey data 
enable us to tackle this problem in a statistical sense, looking for
the general trends that reveal the fundamental processes driving the
formation of galaxies, confronting the observations with models that
encapsulate the main mechanisms transforming gas into stars and
framing this in a cosmological context. Scaling trends such as the 
colour-magnitude relation, the fundamental plane, or the Tully-Fisher
relation provide evidence of a strong trend between a local physical quantity,
say gravitational potential in its many guises, and the formation of
galaxies. For instance, we now know that there is a strong
correlation between the age and the mass of a galaxy, with more
massive systems being preferentially older, and predominantly
supported as hot dynamical systems \citep[see, e.g.,][for a general review]{SM:12}.  The well-established bimodality \citep[e.g.,][]{Strat:01,IB:04,JA:19} reveals a blue cloud/red
sequence transition where the processes that quench star formation appear to
hold the key to understanding how galaxies evolve. In this context, an important 
issue concerns the ``drivers'' of galaxy formation,
i.e. the fundamental properties that control this evolution.
A large number of papers look for these
drivers, mainly proposing options at the galactic level, i.e. scales of several kpc, vs environment,  
group/cluster scales of hundreds of kpc and beyond \citep[to name a few,][]{Weinmann:06,Peng:10,Rogers:10,Etherington:15}.
At the galaxy level, relevant 
parameters are the mass, velocity dispersion, gravitational potential, dynamical support
or morphology. 
The more recent IFU-based surveys \citep[e.g.][]{SAURON:01,ATLAS3D:11,CALIFA:12,SAMI:15,MANGA:15} brought
this analysis into a new level, as galaxies could be probed spectroscopically
in much more detail, regarding both the dynamical state and the stellar
population content, with fundamental results into the way these two aspects
relate to each other \citep[e.g.,][]{Capp:11,Wang:24}.

While many of the IFU-based studies statistically approach the
spatially resolved properties with radial gradients of targeted
observables, such a result only produces a first view of
resolved galaxy formation, and depends on how the local properties
match with galactocentric distance. Gradient studies of stellar
populations reveal interesting trends where the central regions tend
to be dominated by old, metal rich stars, whereas the outer regions tend to
be  metal poorer, reflecting a different formation mode
\citep[see, e.g.][]{HK:10,FLB:12,RGD:15,Greene:15,Parikh:18,Parikh:21,SAMIGrad:19,Zibetti:20}.
These observational trends can be explained within a simple framework based on 
simulations \citep{Oser:10}, where
most of the formation of a galaxy is separated into two major phases,
an in-situ phase that generates the 
stellar component of the most massive, parent, halo, and the ex-situ phase
that consists of later additions to the galaxy from mergers \citep{Naab:09}. While this
working hypothesis is a powerful way to understand the spatially resolved
results, it hardwires the interpretation to monolithic versus hierarchical
growth. 

In a insightful review, \citet{Sanchez:21} suggested that local
properties within galaxies are also subject to the same scaling
relations as those found over galaxy scales, so that the latter can be interpreted as an 
integrated version of the former. Such an interpretation goes back to scaling
relations such as the Schmidt-Kennicutt law that defines the  star formation
rate by the local gas density, either as a
projected surface density \citep{Kenn:89} or the 3D volume density \citep{Schmidt:59}.
Local relations provide a more detailed framework than the standard 
two-phase scenario and give small scale mechanisms a more important
role.  Our paper takes this point further, adopting the ansatz that a
local observable -- roughly defined over a physical scale of
$\sim$2\,kiloparsec (limited by the size of the optical fibre)  -- mostly controls the overall properties of the
stellar populations, therefore also determining its past star formation and chemical
enrichment histories. The excellent quality of the publicly available data from the
SDSS-IV IFU survey MaNGA \citep{Manga} allows us to
tackle such a proposal. 
In addition, we also explore which of the typical local
properties are more strongly correlated, and thus can potentially serve as the
major driver of galaxy evolution. Our conclusions strongly favour
{\sl local} stellar velocity dispersion as the dominant driver. Such
a hypothesis requires a revision of some of the established
ideas regarding galaxy formation, and offers a strong constrain to test the
validity of hydrodynamical cosmological simulations.

This paper is structured as follows. After this introductory section, we
describe the MaNGA IFU dataset in Section~\ref{Sec:Sample}, followed by the
description of the scaling properties in Section~\ref{Sec:Ansatz} that motivate
us to propose the ansatz that the star formation histories are mainly
controlled by physical quantities over sub-galaxy scales, with the
velocity dispersion ($\sigma$) posited as the main driver. Thus far these trends
are shown as observational correlations. In Section~\ref{Sec:Driver} we further
consider whether $\sigma$ and not stellar mass represents the cause of
these trends. Section~\ref{Sec:BCGVRS} takes a look at the connection between the
observed local trends and the evolutionary stage of the galaxies. Finally, we
summarise our conclusions in Section~\ref{Sec:Conc}.

\begin{figure}
\includegraphics[width=80mm]{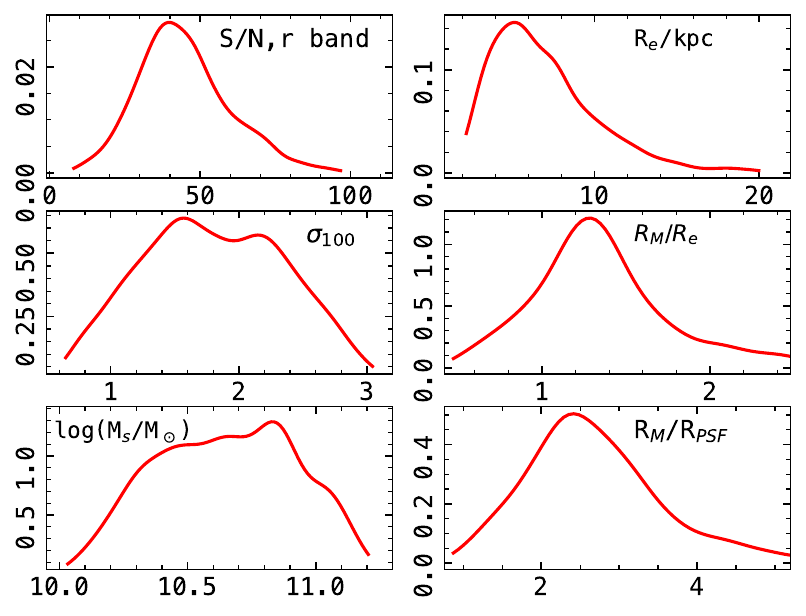}
\caption{Distribution of general observables of the galaxy sample
  from SDSS-MaNGA.  The corresponding parameter (horizontal axis) is
  labelled in each panel. Clockwise from top-left we show the signal
  to noise ratio of the spectra within one effective radius in the
  SDSS-$r$ band, the effective radius in physical units, the ratio
  between the galactocentric distance of the outermost useable spaxel
  in each galaxy (R$_M$) and the effective radius, the ratio between
  R$_M$ and R$_{\rm PSF}$ (i.e. the HWHM of the point spread function), the (logarithm) of
  the stellar mass in solar units, and the averaged stellar velocity
  dispersion in units of 100\,km\,s$^{-1}$.}
\label{fig:hist}
\end{figure}

\begin{figure*}
\includegraphics[width=160mm]{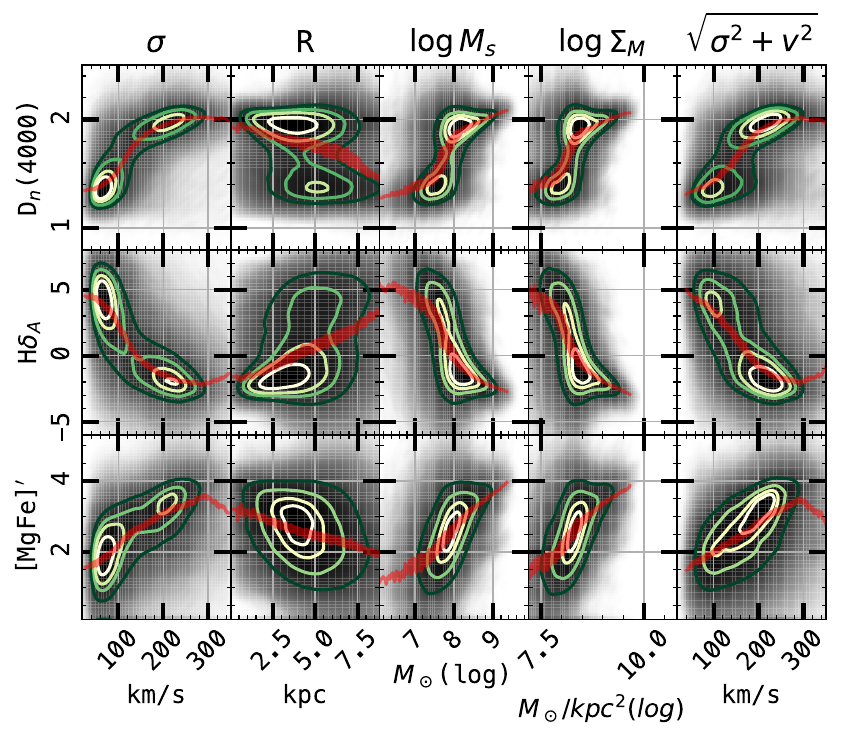}
\caption{Bivariate plots of three fundamental line strengths shown with respect to five indicators of galaxy ``position'', as defined {\sl for each individual spaxel}, (from left to right) velocity dispersion, galactocentric distance, stellar mass, surface stellar mass density, and a proxy of kinetic energy, $\sqrt{\sigma^2+v^2}$. The data consists of over 1 million measurements comprising {\sl individual} spaxels, regardless of the galaxies they correspond to. The contours engulf (from the inside out) 25, 50, 75 and 90\% of all data points, starting in the regions of higher data point density. The red line in each panel follows the running median of the distribution with respect to the abscissa.}
\label{fig:spxALLAbs}
\end{figure*}

\section{The sample}
\label{Sec:Sample}

The parent catalogue for this work is the \citet{JA:19,JA:20}
sample that is based on a detailed analysis of high quality data from
the Legacy SDSS spectra, classified into the three standard
evolutionary states (blue cloud, green valley and red sequence) using
one of the most robust indicators of the stellar population content,
the 4000\AA\ break strength. That sample is restricted in redshift
($0.05\leq z\leq 0.1$) and in the signal to noise ratio of the single fibre spectra
(median S/N$>$10 over pixels in the SDSS-$r$ band), and consists of over 200
thousand galaxies.  We cross-match this catalogue with the DR17
version \citep{DR17} of the SDSS-IV/MaNGA \citep{Manga} IFU survey,
that comprises $\sim$10,000 galaxies. We also make use of the Marvin
\citep{Marvin} data products. The cross-match results in a
set of 2,034 galaxies, but one galaxy does not have the
science-ready data from Marvin (mangaID 1-80510), and a few other galaxies have repeated
observations (see Table~1 from \citealt{DAP}), for which we retrieve the ones with
the best seeing according to the SEEMED keyword. The final set comprises 2,024
galaxies, and constitutes our working sample.
Fig.~\ref{fig:hist} shows the distribution of the sample with respect
to a number of important parameters defined in each galaxy observation (clock-wise from top-left:
signal-to-noise ratio measured within one effective radius, the
physical effective radius in kpc -- assuming a vanilla flavoured
$\Lambda$CDM cosmology with $h$=0.7 and $\Omega_m$=0.3 -- the
galactocentric distance of the outermost useable spaxel (R$_M$) in units of
$R_e$, the same R$_M$ in units of the HWHM of the PSF (defined as R$_{\rm PSF}$),
stellar mass, and velocity
dispersion in units of 100\,km\,s$^{-1}$. By ``useable'' spaxels, we
enforce a minimum S/N (also in the SDSS-$r$ band) of 5 for {\sl
  individual spectra}. In addition, we discard
spaxels where the individual velocity dispersion estimate is uncertain,
or outside of the [50,300]\,km/s interval. Both constraints ensure that
the resulting individual measurements are reliable. The final sample of
2,024 galaxies produces a set of 1,025,841  spaxels.

\begin{figure*}
\includegraphics[width=160mm]{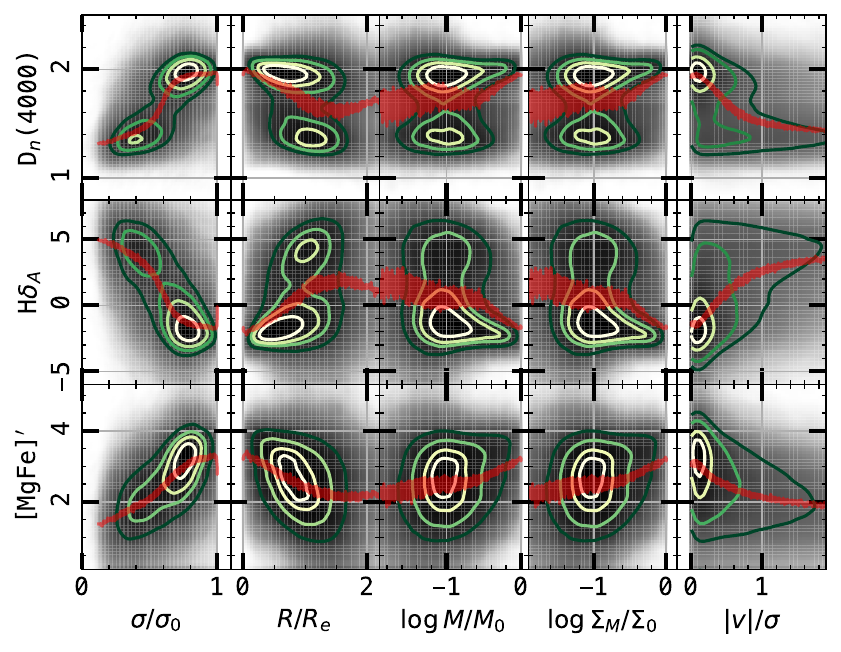}
\caption{Equivalent of Fig.~\ref{fig:spxALLAbs} for relative
  observables, as described in the text. In addition to (from left to
  right) stellar velocity dispersion, galactocentric distance, stellar
  mass and stellar mass surface density, we also include the ratio
  between the absolute value of bulk velocity and velocity dispersion
  of the stellar component, as an estimator of rotational support. All
  measures are referred to over 1 million individual spaxels. The contours engulf
  (from the inside out) 25, 50, 75 and 90\% of all data points,
  starting in the regions of higher data point density. The red line in each panel tracks the running median of the distribution with respect to the abscissa.}
\label{fig:spxALLRel}
\end{figure*}

The novel approach of this paper lies in our relinquishing the connection of the data with the ``host'' 
galaxy, turning instead to the analysis of {\sl individual
  spaxels}, representing local regions of galaxy formation and
evolution. In other words, we consider this sample as a
set of over one million good quality independent spectra of local star forming regions.
At the median redshift of the sample (z$_M$=0.065), the size of a spaxel fibre
(2\,arcsec diameter) maps a physical scale of 2.5\,kpc.  
We explore several types of ``local indicators'': stellar velocity
dispersion ($\sigma$, measured locally in each spaxel), projected 
galactocentric distance ($R$), stellar mass ($M_s$), stellar mass surface density
($\Sigma_{M}$, defined as the ratio between the stellar mass in the spaxel and
the physical area spanned by the fibre at the location of the galaxy) and
the bulk velocity ($v$). These indicators can be defined in an
absolute sense (i.e. each spaxel is labelled by a $\sigma$ in km/s, 
an $R$ in kpc, a $M_s$ in $M_\odot$, a $\Sigma_{M}$ in $M_\odot$\,kpc$^{-2}$), 
or as a relative dimensionless quantity: $\sigma, M_s$ and $\Sigma_{M}$ being
given as a fraction of the maximum value of the relevant quantity in the 
corresponding galaxy, and $R$ measured in units of the effective
radius, $R_e$. For the bulk velocity, we use two definitions: the absolute measurement
combines it with velocity dispersion, $\sqrt{\sigma^2+v^2}$ and serves as a proxy of
kinetic energy. The relative definition is the standard ratio $|v|/\sigma$ 
that traces the level of rotational support versus random motion,
a direct indicator of morphology \citep[see, e.g.,][]{Capp:11}.
In Appendix~\ref{app:single}, we show the distribution of spaxel measurements for a
few individual galaxies, comparing them with respect to the
distribution of the complete spaxel sample as we will show in the next section.
We emphasize that our analysis rests on the assumption that the statistical distribution
of individual spaxels reflects an inherent causality at sub-galaxy ($\sim$2\,kpc) scales,
regardless of the stellar mass, morphology, etc, corresponding to the parent galaxy.

\begin{table*}
  \centering
  \caption{Correlation coefficients between stellar population line strengths and observables measured in an absolute way in spaxels with S/N$\geq$5 (see fig~\ref{fig:spxALLAbs})}
  \begin{tabular}{cccccc} 
    \hline
    Index & $\sigma$ & $R$ & $\log M_s$ & $\log\Sigma_M$ & $\sqrt{\sigma^2+v^2}$\\
    \hline
    \multicolumn{6}{c}{Correlation coefficient}\\
    \hline
    D$_n$(4000)    & $+$0.756$\pm$0.001 & $-$0.236$\pm$0.001 & $+$0.578$\pm$0.001 & $+$0.548$\pm$0.001 & $+$0.610$\pm$0.001\\
    H$\delta_A$    & $-$0.669$\pm$0.001 & $+$0.297$\pm$0.001 & $-$0.562$\pm$0.001 & $-$0.563$\pm$0.001 & $-$0.522$\pm$0.001\\
    $[{\rm MgFe}]^\prime$ & $+$0.548$\pm$0.001 & $-$0.310$\pm$0.001 & $+$0.518$\pm$0.001 & $+$0.511$\pm$0.001 & $+$0.401$\pm$0.001\\
    $\log t_{LW}$ & $+$0.361$\pm$0.001 & $-$0.162$\pm$0.001 & $+$0.614$\pm$0.001 & $+$0.645$\pm$0.001 & $+$0.275$\pm$0.001\\
    $\log t_{MW}$ & $+$0.183$\pm$0.001 & $-$0.045$\pm$0.001 & $+$0.434$\pm$0.001 & $+$0.466$\pm$0.001 & $+$0.134$\pm$0.001\\
    $\log Z_{LW}$ & $+$0.490$\pm$0.001 & $-$0.198$\pm$0.001 & $+$0.476$\pm$0.001 & $+$0.454$\pm$0.001 & $+$0.452$\pm$0.001\\
    $\log Z_{MW}$ & $+$0.403$\pm$0.001 & $-$0.144$\pm$0.001 & $+$0.454$\pm$0.001 & $+$0.433$\pm$0.001 & $+$0.350$\pm$0.001\\
    \hline
    \multicolumn{6}{c}{Standard deviation (and 1$\sigma$ error)}\\
    \hline
    D$_n$(4000)    & 0.169$\pm$0.037 & 0.273$\pm$0.019 & 0.223$\pm$0.050 & 0.228$\pm$0.050 & 0.213$\pm$0.027\\
    H$\delta_A$    & 2.139$\pm$0.328 & 2.888$\pm$0.289 & 2.428$\pm$0.577 & 2.413$\pm$0.597 & 2.516$\pm$0.244\\
    $[{\rm MgFe}]^\prime$ & 0.818$\pm$0.042 & 0.930$\pm$0.122 & 0.828$\pm$0.124 & 0.830$\pm$0.139 & 0.881$\pm$0.047\\
    $\log t_{LW}$ & 0.234$\pm$0.025 & 0.252$\pm$0.038 & 0.188$\pm$0.037 & 0.179$\pm$0.033 & 0.243$\pm$0.016\\
    $\log t_{MW}$ & 0.216$\pm$0.030 & 0.221$\pm$0.030 & 0.188$\pm$0.041 & 0.182$\pm$0.039 & 0.219$\pm$0.021\\
    $\log Z_{LW}$ & 0.142$\pm$0.026 & 0.165$\pm$0.016 & 0.142$\pm$0.034 & 0.145$\pm$0.033 & 0.147$\pm$0.020\\
    $\log Z_{MW}$ & 0.235$\pm$0.070 & 0.266$\pm$0.039 & 0.224$\pm$0.074 & 0.226$\pm$0.071 & 0.242$\pm$0.058\\
    \hline
  \end{tabular}
  \label{tab:corrAbs}
\end{table*}

\begin{table*}
  \centering
  \caption{Correlation coefficients between stellar population line strengths and observables measured in a relative way in spaxels with S/N$\geq$5 (see fig~\ref{fig:spxALLRel})}
  \begin{tabular}{cccccc} 
    \hline
    Index & $\sigma/\sigma_0$ & $R/R_e$ & $\log M_s/M_0$ & $\log\Sigma_M/\Sigma_0$ & $|v|/\sigma$\\
    \hline
    \multicolumn{6}{c}{Correlation coefficient}\\
    \hline
    D$_n$(4000)    & $+$0.651$\pm$0.001 & $-$0.207$\pm$0.001 & $+$0.062$\pm$0.001 & $+$0.060$\pm$0.001 & $+$0.036$\pm$0.001\\
    H$\delta_A$    & $-$0.605$\pm$0.001 & $+$0.240$\pm$0.001 & $-$0.117$\pm$0.001 & $-$0.117$\pm$0.001 & $-$0.007$\pm$0.001\\
    $[{\rm MgFe}]^\prime$ & $+$0.520$\pm$0.001 & $-$0.292$\pm$0.001 & $+$0.175$\pm$0.001 & $+$0.177$\pm$0.001 & $-$0.027$\pm$0.002\\
    $\log t_{LW}$ & $+$0.317$\pm$0.001 & $-$0.111$\pm$0.001 & $+$0.398$\pm$0.001 & $+$0.396$\pm$0.001 & $-$0.004$\pm$0.002\\
    $\log t_{MW}$ & $+$0.136$\pm$0.001 & $+$0.004$\pm$0.001 & $+$0.318$\pm$0.001 & $+$0.315$\pm$0.001 & $+$0.007$\pm$0.002\\
    $\log Z_{LW}$ & $+$0.458$\pm$0.001 & $-$0.207$\pm$0.001 & $+$0.118$\pm$0.001 & $+$0.120$\pm$0.001 & $+$0.013$\pm$0.001\\
    $\log Z_{MW}$ & $+$0.385$\pm$0.001 & $-$0.139$\pm$0.001 & $+$0.125$\pm$0.001 & $+$0.124$\pm$0.001 & $+$0.013$\pm$0.002\\
    \hline
    \multicolumn{6}{c}{Standard deviation (and 1$\sigma$ error)}\\
    \hline
    D$_n$(4000)    & 0.208$\pm$0.030 & 0.274$\pm$0.021 & 0.280$\pm$0.015 & 0.280$\pm$0.015 & 0.234$\pm$0.029\\
    H$\delta_A$    & 2.370$\pm$0.285 & 2.912$\pm$0.347 & 3.001$\pm$0.266 & 3.001$\pm$0.266 & 2.614$\pm$0.257\\
    $[{\rm MgFe}]^\prime$ & 0.839$\pm$0.035 & 0.933$\pm$0.115 & 0.963$\pm$0.106 & 0.963$\pm$0.106 & 0.889$\pm$0.047\\
    $\log t_{LW}$ & 0.240$\pm$0.026 & 0.254$\pm$0.038 & 0.221$\pm$0.036 & 0.221$\pm$0.036 & 0.244$\pm$0.020\\
    $\log t_{MW}$ & 0.218$\pm$0.028 & 0.222$\pm$0.026 & 0.199$\pm$0.032 & 0.199$\pm$0.032 & 0.218$\pm$0.025\\
    $\log Z_{LW}$ & 0.147$\pm$0.023 & 0.163$\pm$0.021 & 0.167$\pm$0.015 & 0.167$\pm$0.015 & 0.159$\pm$0.016\\
    $\log Z_{MW}$ & 0.241$\pm$0.065 & 0.265$\pm$0.048 & 0.268$\pm$0.031 & 0.268$\pm$0.031 & 0.258$\pm$0.046\\
    \hline
  \end{tabular}
  \label{tab:corrRel}
\end{table*}

\section{The ansatz: Local properties drive the average star formation history}
\label{Sec:Ansatz}

Fig.~\ref{fig:spxALLAbs} shows, as a density plot,  the distribution of spaxel data of the whole sample for three
targeted line strengths, from top to bottom, the 4000\AA\ break strength, adopting the
definition of \citet{Balogh:99}, the wide (A) definition of the H$\delta$ Balmer line of \cite{WO:97}, and the [MgFe]$^\prime$ index that combines the
traditional Lick indices Mgb, Fe5270 and Fe5335 in a way that minimises the dependence
on the [Mg/Fe] abundance ratio \citep{TMB:03}. These three indices lock a large amount of ``information''
(in the entropy sense) or variance, so they are ideal indicators to explore the
population properties, especially in data covering a wide range of S/N values \citep[see][]{Entropy}.
Very roughly, we can assume that D$_n$(4000) traces overall stellar
age, H$\delta_A$ is prominent in regions with recent star formation activity, and
[MgFe]$^\prime$ features a strong dependence with metallicity. However, the reader should
be cautious in the interpretation as these indices (or any other) have a substantial
age-metallicity degeneracy \citep[e.g.][]{ameg,FLB:13}, and are strongly
correlated with respect to the fundamental population properties \citep{Entropy}. For reference, we also 
include below the analysis for the stellar age and metallicity derived by adopting a standard
methodology \citep{FireFly}.

Fig.~\ref{fig:spxALLAbs} presents the data with respect to the indicators
defined in an absolute sense: stellar velocity dispersion ($\sigma$, in km/s),
galactocentric distance ($R$, in kpc), stellar mass ($M_s$, in $M_\odot$), 
stellar mass surface density ($\Sigma_M$ in $M_\odot$\,kpc$^{-2}$), and
a proxy for the kinetic energy, i.e. combining velocity dispersion and
bulk motion in quadrature ($\sqrt{\sigma^2+v^2}$ in km/s). We emphasize
that these are local indicators, defined for specific spaxels, and all
the spaxels in all galaxies from the sample are included here to
produce these distributions. In addition to the greyscale density
plot, to guide the eye, we overlay contours at levels that engulf (from
the inside out) 25, 50, 75 and 90\% of the total set.  Each of the
indices is corrected for velocity dispersion effects using the term
provided in the Marvin dataset (for instance: specindex\_corr\_hdeltaa
for the correction of the H$\delta_A$ line strength, defined as a
multiplicative correction). Our hypothesis -- that the stellar population content is ``driven'' by
local quantities -- is supported by the trends shown in the figure,
where the line strengths are strongly correlated with {\sl local}
velocity dispersion, regardless of galaxy type, mass, etc. In addition,
Fig.~\ref{fig:spxALLRel} shows the same observables defined in the previous
figure, but in a relative sense: for $\sigma$, $M_s$ and $\Sigma_M$ we take
the ratio between the measured value in each spaxel and the maximum of the
distribution {\sl for each individual galaxy}, and for $R$ it is given as the
ratio with respect to the effective radius, taken from the official SDSS/MaNGA
data. Finally, we include the ratio between the
bulk velocity in absolute value and the velocity dispersion (both concerning
the stellar component).

\begin{figure*}
\includegraphics[width=160mm]{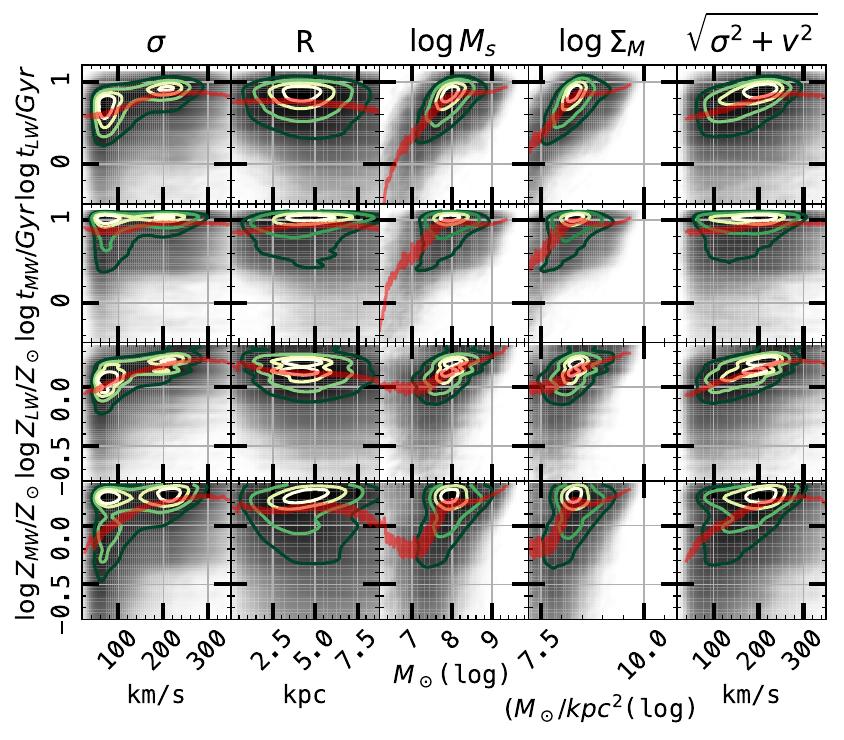}
\caption{Equivalent of Fig.~\ref{fig:spxALLAbs} for the Luminosity- (LW)
  and Mass-Weighted (MW) stellar age and metallicity, as derived from the
  FireFly fitting procedure \citep{FireFly}.}
\label{fig:spxALLAbsAge}
\end{figure*}

A first test to determine which of these local indicators is dominant, we
calculate the Pearson correlation coefficient (r$_{\rm XY}$) for all the bivariate
plots presented in Fig.~\ref{fig:spxALLAbs} 
for absolute observables and Fig.~\ref{fig:spxALLRel}
for relative observables. The results are shown in the top
portions of Tabs.~\ref{tab:corrAbs} and \ref{tab:corrRel}, respectively.
The quoted uncertainties correspond to 1$\sigma$ of the distribution
of correlation coefficients derived from 1,000 randomly selected samples 
each one comprising 75\% of the whole set. 
This coefficient is defined to look for linear
trends in data sets. Figs.~\ref{fig:spxALLAbs} and \ref{fig:spxALLRel}
show that these correlations are not necessarily linear, but r$_{\rm XY}$
provides a useful quantification of correlatedness -- much in the same
way as covariance in non-Gaussian distributions. We also explored non-linear
statistics, such as the Spearman correlation coefficient (r$_{\rm S}$), which is better
suited to track monotonicity. We find very similar results that confirms these results.
For instance, the correlation between (absolute) velocity dispersion and
D$_n(4000)$ is r$_{\rm XY}$=$+$0.756, r$_{\rm S}$=0.774, and the equivalent result for the trend with (absolute) galactocentric radial distance and D$_n$(4000) is r$_{\rm XY}$=$-$0.236, r$_{\rm S}$=$-$0.224.

A second measure of the correlation, better suited to non-linear trends,
say between X=$\sigma$ and Y=$D_n(4000)$, 
involves producing a running median $Y_R=f(X)$ by taking sliding intervals
(of size 1,000) in increasing X. For each interval we define
the general trend by the running median: $f(X_i) = {\rm median}(X_i)$, shown as red lines in each panel
of Figs.~\ref{fig:spxALLAbs} and \ref{fig:spxALLRel}).
We now define the residuals with respect to this trend: $\delta_i\equiv (Y_i - f(X_i))$
for each data point (i.e. spaxel), from which
we quote the standard deviation: 
$\sqrt{\langle\langle\delta^2\rangle - \langle\delta\rangle^2\rangle}$, 
in the bottom portion of Tabs.~\ref{tab:corrAbs} and \ref{tab:corrRel}, along with its
own standard deviation, serving as an indicator of the quality of this number as a tracer
of correlation.

The strongest correlation is found for stellar (local) velocity dispersion in all
three line strengths, especially D$_n$(4000). Note that of the three
choices, the 4000\AA\ break strength is the one least sensitive to
systematic effects of velocity dispersion\footnote{because it is defined in a wider spectral range.},
confirming the strong
relation between stellar population content and $\sigma$,
i.e. excluding a potential systematic. Both the absolute case
($\sigma$ presented in km/s) or the relative one (adopting the
dimensionless ratio $\sigma/\sigma_0$) produce similar 
coefficients, although the latter appears less correlated with
population properties. The kinetic energy proxy ($\sqrt{\sigma^2+v^2}$) is
also strongly correlated, but less so than velocity dispersion.
Moreover, note the very weak correlation of $|v|/\sigma$, which leads
us to conclude that the strong correlation found in $\sqrt{\sigma^2+v^2}$ is
dominated by $\sigma$. It is also worth noting that the measured bulk velocity
depends on inclination, weakening the expected trends. Regarding stellar
mass and stellar mass surface density, it is quite remarkable to note
the strong correlation for the absolute estimators
(Tab.~\ref{tab:corrAbs}) but the substantially lower level of
correlation for the relative observables (Tab.~\ref{tab:corrRel}). We
emphasize that the relative values are simply referenced with respect
to the maximum value of the observable in each galaxy. Such a
behaviour is not found for velocity dispersion where both absolute and
relative estimates appear strongly correlated with the line strengths.

In stark contrast, galactocentric distance is very weakly correlated with any of the
population parameters, either in absolute or relative estimates.
Our measurements of galactocentric distance
adopt the official MaNGA parameter based on the elliptical Petrosian
half-light radius as R$_e$, a more robust indicator than parameters
based on S\'ersic fitting \citep{Wake:17}.  Note also the jagged
behaviour of the running median lines, especially concerning radial
distance, produced because of two factors: weaker correlation and
in some cases, the presence of a bimodality of
line strength measurements within the same range of
galactocentric distance. In contrast, the plots with velocity dispersion separate better
the two distributions (e.g. strong vs weak 4000\AA\ break strength).
We note that while spatial resolution may introduce a systematic,
mainly as the size of the PSF is not small with respect to the extent
of the observations (see Fig.~\ref{fig:hist}), the effect would induce lower correlations in
both galacto-centric distance and velocity dispersion, whereas we find
a notable difference between these two local indicators.

In addition to the line strengths directly measured in the spaxels, we
include in Fig.~\ref{fig:spxALLAbsAge} the distribution of population
parameters derived from the spectra. From bottom to top we show the
stellar age (luminosity weighted and mass weighted) and the
metallicity (also luminosity weighted and mass weighted). The results
are retrieved from the Portsmouth port of the MaNGA database using the
Firefly code \citep{FireFly}. These parameters are now subject to the
systematics produced by fitting the spectra to a set of population
synthesis models. The same measurements of the correlation coefficient and
scatter with respect to the running median is shown for these parameters in
Tabs.~\ref{tab:corrAbs} for absolute indicators and \ref{tab:corrRel} for
relative indicators.
We should emphasize that the stellar mass used in this work is also
dependent on the same type of analysis -- in order to translate the observed 
flux into a stellar mass. Therefore, it comes as no suprise that these parameters
show substantial correlation. However stellar velocity dispersion -- a more
independent observable to the derivation of population parameters -- is also
found to correlate at a similar level.

\begin{figure}
\includegraphics[width=80mm]{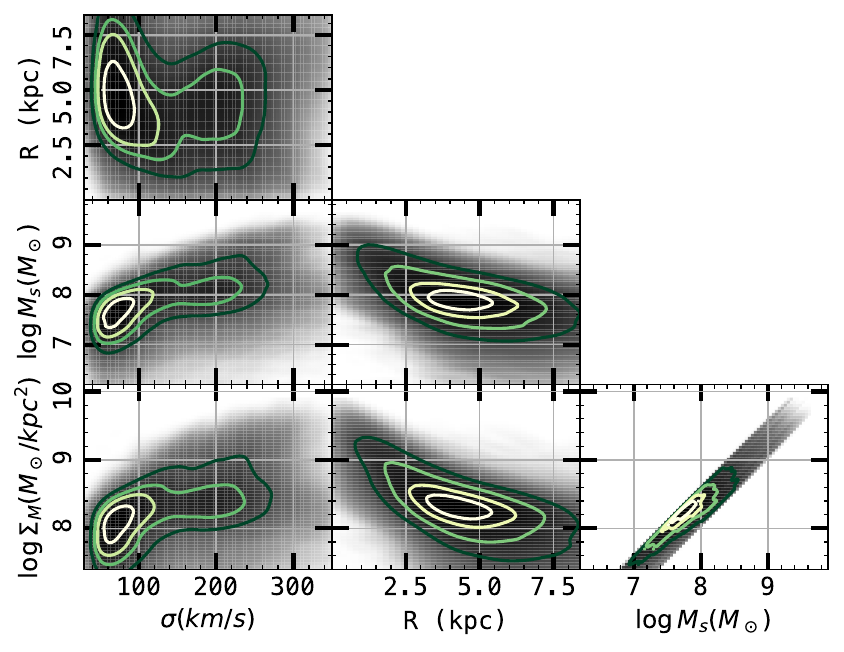}
\caption{Relationship between local indicators, shown as a density plot of the distribution of all spaxel data. The contours engulf (from the inside out) 25, 50, 75 and 90\% of all data points, starting from the regions of higher data point density.}
\label{fig:LocInd}
\end{figure}

\section{Which one is the major driver of population properties?}
\label{Sec:Driver}

Finding the fundamental driver of star formation in galaxies is a
complicated task.  Most of the work in statistics is geared towards
the search, quantification and analysis of correlations, but the
mantra of causal inference states that ``correlation does not imply
causation\footnote{but sometimes it does!}''. Indeed, interesting
cases abound where highly correlated variables clearly do not cause
the trend under study -- such as the textbook example of the
correlation between ice cream sales and drowning, where the
fundamental ``driver'' of the relation is climate (typically called
confounder variable, see, e.g. \citealt{BookofWhy}).  An illustrative
point regarding the relationship of the local indicators explored in
this paper is shown in Fig.~\ref{fig:LocInd}, where the four choices
are compared against one another as a density plot. The most conspicuous
relation is unsurprisingly found between stellar mass and surface stellar
mass density, as both use M$_s$ and $\Sigma_M$ simply adds a second variable,
$R$, that also correlates with stellar mass. A very relevant plot is the
weak correlation between $\sigma$ and $R$, that, linked to the previously
found correlation with line strengths, leads us to discard galactocentric
distance as a key variable in the analysis. Stellar mass and $\sigma$ do have
a substantial correlation, and it is worth noting that these two variables are
derived in independent ways, so this trend shows a physical connection. We 
therefore consider $\sigma$ and $M_s$ as candidates of the ``main'' variable. But
which of these two is more important?

\begin{figure}
\includegraphics[width=80mm]{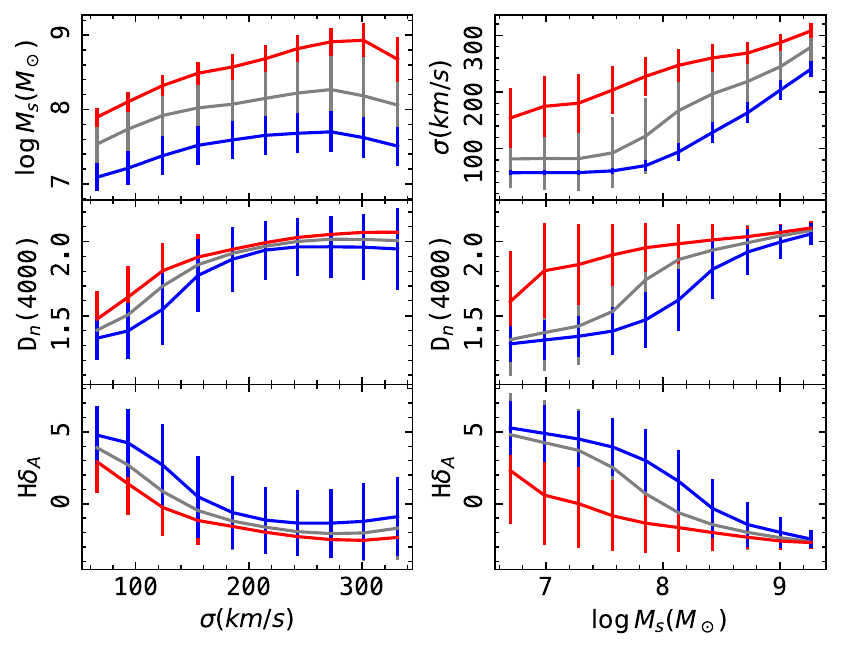}
\caption{Evolution of the median of 4000\AA\ break strength (middle) and H$\delta_A$ (bottom) as a function of velocity dispersion (left) or stellar mass (right). The whole sample is shown in grey, along with the subsets at low (25 percentile, blue) and high (75 percentile, red) of the distribution within each bin in $\sigma$ or $\log M_s$. The points are the median values, whereas the error bars represent the standard deviation in each bin. The top panels illustrate the sample selection strategy.}
\label{fig:test_sigMass}
\end{figure}

A first test is shown in Fig.~\ref{fig:test_sigMass}, where the sample is split
primarily with respect to velocity dispersion (left column), or stellar mass
(right column). In each case, we bin the main variable in ten intervals within which we
assume the main variable is roughly ``fixed'', i.e. defined by the chosen bin. In each interval, the other variable
($\log\,$M$_s$ when binning in $\sigma$ and vice-versa) is used to select a subset at the
25th and 75th percentile levels from the distribution. The top panels in each case trivially show the sample selection, 
with the high (low) subsets shown in red (blue),
and the total sample in grey. The points represent the median in each case, along with
the standard deviation, shown as an error bar. The middle and bottom panels show the
equivalent median and standard deviation of the 4000\AA\ break and the H$\delta_A$ line strength.
The variation in the line strengths at fixed $\sigma$ is smaller than the corresponding one at
fixed stellar mass, suggesting that stellar mass plays some role, but in a subdominant way with
respect to velocity dispersion.

\begin{figure}
\includegraphics[width=80mm]{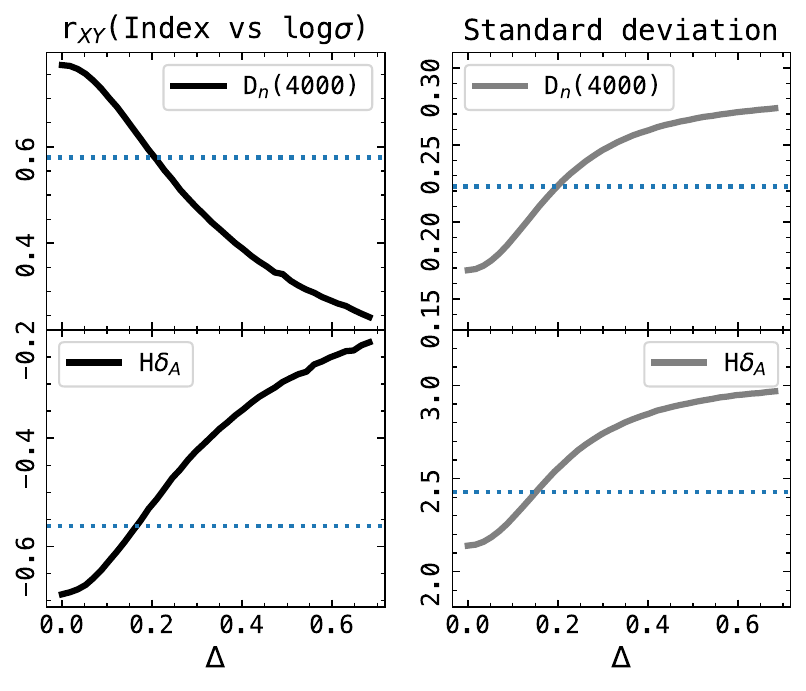}
\caption{The correlation coefficient (left) and standard deviation of the residual (right) is shown as Gaussian noise is randomly added to the measurement of stellar velocity dispersion. Two line strengths are considered: 4000\AA\ break (top) and H$\delta_A$ (bottom). As the noise level ($\Delta$) increases, the correlation (standard deviation) decreases (increases). For reference, the values obtained for the relations with respect to stellar mass are shown as blue horizontal dotted lines.}
\label{fig:corr_noise}
\end{figure}

Previous studies aiming at the cause of the observed population trends
within galaxies only conclude that stellar velocity dispersion is the
most correlated variable with stellar populations \citep[see,
  e.g.,][]{SAMIGrad:19}. However, one could argue that $\sigma$ is the
measurement with the lowest intrinsic uncertainty -- its derivation simply depends
on a comparison of many absorption lines in the spectrum of the galaxy
with a smoothing kernel, and thus minimises the potential systematics
regarding the method. In contrast, stellar mass suffers from a more
elaborate set of systematic effects that depend on the modelling of
the stellar populations to infer a mass-to-light ratio. We
reject radial distance as a major driver, as it
suffers from relatively low measurement systematics, and the
correlation coefficients are substantially lower than the other local
measures (while correlation is not causation, we need a strong
correlation to justify the cause). Fig.~\ref{fig:corr_noise} compares
the Pearson correlation coefficient and associated scatter of the
running median (as presented in Tab.~\ref{tab:corrAbs}) of the trends
between $D_n(4000)$ (top) or H$\delta_A$ (bottom) and stellar velocity
dispersion. The horizontal axis corresponds to a parameter, $\Delta$
that represents the standard deviation of an extra component of random Gaussian
noise added to each measure of velocity dispersion. As $\Delta$
increases, we expect the correlation to decrease. Such an experiment
seeks to answer the following question: what type of additional random
noise is needed to make the correlation of the population indices with
velocity dispersion comparable to that of the trends with stellar
mass? For reference, the latter are marked with a horizontal dashed
line. In all cases we find that Gaussian noise at the level
$\Delta\sim 0.2$\,dex can mimic the trends found. It is not uncommon
to expect such an uncertainty in stellar mass estimates, although
comparisons of independent methods suggest lower values, around $\lesssim 0.1$\,dex
\citep[see, e.g.][]{Santini:15, Pacifici:23, Dogruel:23}. These uncertainties
are mainly produced by the methodology, regarding template fitting,
mismatch of targeted populations, photometric error, fitting of the total
luminosity, etc. The additional error we need in velocity dispersion is
on the high side to bring the observed correlations with $\sigma$ and
M$_s$ in line. Therefore we
conclude that it may be possible to confirm $\sigma$ as the major
driver of population properties, but unknown uncertainties in the
stellar mass may also make this parameter a major driver.
In any case, note that while M$_s$ and $\sigma$ are found
in very different ways, the trends are fully consistent with the
ansatz that local variables determine, statistically, the stellar
population properties in galaxies.

\section{A note on the link with the evolutionary stage}
\label{Sec:BCGVRS}

Once we determine the potential main driver of population properties over small scales, a
question remains regarding the connection with the general evolutionary stage, as
posed by the well-established bimodality in colour \citep{Strateva:01}, line strengths \citep{JA:19} and
star formation rate \citep{Speagle:14}. A clear separation is found with regards to star formation activity,
or lack thereof, along with a intriguing transitioning phase, the Green Valley, that encode
the details of the feedback mechanisms that quench galaxy growth \citep[see, e.g.][]{Schaw:14,Salim:14}.

In this exercise, we focus on this evolutionary trend by adopting two
classification schemes: one is the segregation into blue cloud (BC),
green valley (GV) and red sequence (RS), as presented in
\citet{JA:19}. In that paper, the selection is based on a bivariate
plot formed by D$_n$(4000) and velocity dispersion, measured in the
SDSS Legacy spectra, measured in single 3\,arsec diameter fibres.  A
second scheme is based on the traditional BPT \citep{BPT} diagram. Using
once more the analysis from the SDSS Legacy (single fibre) spectra,
as presented in the data products of \citet{Jarle:04}, 
we separate galaxies into quiescent (Q), star forming (SF) and AGN.

\begin{figure*}
\includegraphics[width=78mm]{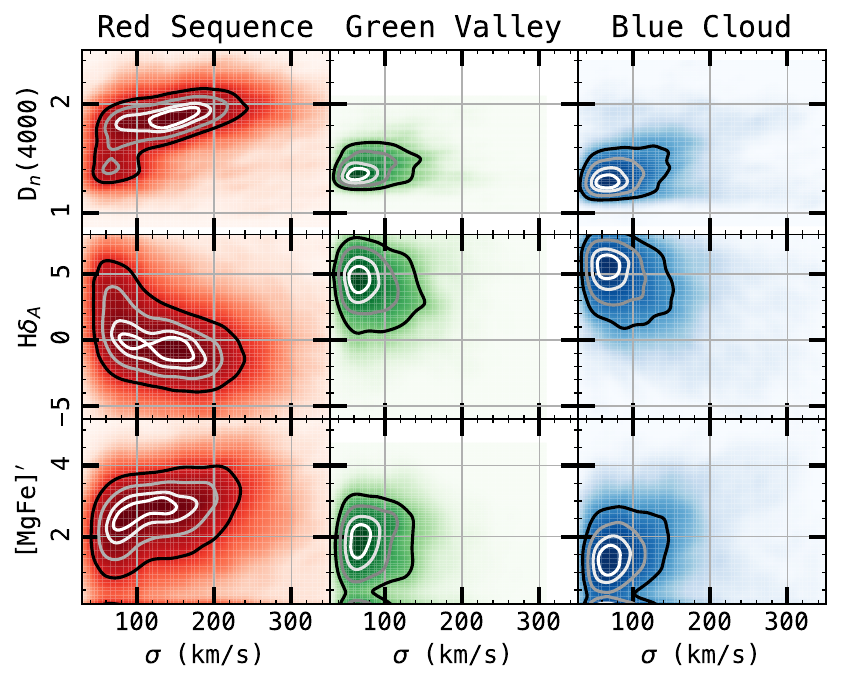}
\includegraphics[width=78mm]{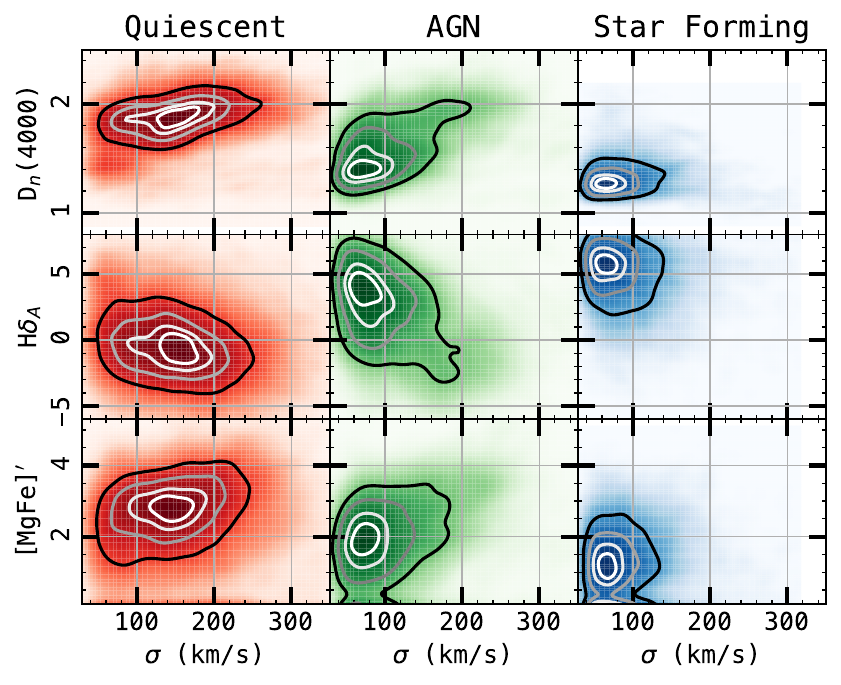}
\caption{Line strength plot for galaxies segregated with respect to the
  classification of \citet{JA:19} ({\sl Left}) or regarding nebular activity
  via the standard BPT diagram ({\sl Right}). The data points correspond to
  individual spaxels in the outermost regions of the chosen galaxies (>1.5Re, S/N>5).}
\label{fig:ProbPlot}
\end{figure*}

Fig.~\ref{fig:ProbPlot} shows the distribution of spaxel data with
respect to $\sigma$ for the three line strengths targeted in this
paper. However, given that the classification method is based on the
central spectra of galaxies, we restrict the sample to the outermost
spaxels (R$>$1.5R$_e$). This allows us to explore the local properties
of individual spaxels without introducing the obvious selection bias. The figure 
complements our previous results as the correlation remains in
all cases. Unsurprisingly, the RS and Q data favours higher
4000\AA\ break strengths and weaker Balmer absorption, along with
higher [MgFe]$^\prime$. The BC and SF sets mostly populate the regions
expected by younger populations, whereas the GV and AGN live somewhere in between
the two major cases of the bimodality. While this is an expected
result, we should emphasize that these data correspond to spaxels
outside of the regions where the selection was produced. Therefore,
while we claim that sub-galaxy (around 2\,kpc) scales drive the trends in the
line strengths, the evolutionary stage (which is defined over galaxy
scales) does indeed affect the distributions. In other words, the
outside spaxels ``know'' they belong to a galaxy of a given
evolutionary type.

Note also the interesting differences between both classification
schemes (left vs right in Fig.~\ref{fig:ProbPlot}). The RS sample does
feature a substantial fraction of weak D$_n$(4000) data at low
$\sigma$, in contrast with the Q sample.  Similarly, the AGN vs GV
comparison also suggests a fraction of AGN that would not be
classified as GV.  Note that the BPT selection relates to
substantially shorter timescales (regarding the nebular emission of
the diffuse gas), whereas the \citet{JA:19} classification is based on
the 4000\AA\ break, therefore associated to the longer timescales of stellar evolution. The consistency confirms that the recent behaviour is statistically
connected with those larger timescales.

\section{Conclusions}
\label{Sec:Conc}

In this paper we make use of the SDSS-IV MaNGA IFU dataset \citep{Manga} in a novel 
way: instead of exploring the data with galaxies as
fundamental units, we adopt the ansatz that local properties --
defined within each spaxel -- determine the stellar population
properties in a statistical way. Given that these properties relate to
the star formation history, this claim goes beyond the standard
framework, and provides an observational trend that any model of galaxy
formation should be able to reproduce. Our sample involves 2,024 high
quality IFU datacubes of galaxies at redshift $0.05<z<0.1$ (median z$_M$=0.065) from which we
extract over one million spaxels, treated independently in this work,
only to be related to quantities defined by the spaxels themselves.
 The analysis makes use of the Marvin data products \citep{Marvin}.

From the set of local properties considered, we find that local
stellar velocity dispersion is the one with the strongest correlation,
although the intrinsically low systematics may be the reason that it
fares better than stellar mass of stellar mass surface
density. Galactocentric radial distance is readily ruled out as a fundamental
driver, as it consistently shows lower correlation coefficients and
higher scatter. In light of this, one should consider that radial
gradients of galaxy properties are a consequence of radial variations in the local 
stellar velocity dispersion at different radii, a tracer of the gravitational
potential. We also note that the dominant role of stellar velocity dispersion
found in these data may parallel the strong trend  between the central
supermassive black hole mass and the velocity dispersion of the bulge \citep{FM:00,KH:13}.
However, we should emphasize that the strong correlation found in our data extends to the
spaxels in the outermost regions (see Fig.~\ref{fig:ProbPlot}), where the potential
contamination from the bulge should be minimal, as illustrated in App.~\ref{app:outer}.

Our results support and complement the suggestion of \citet{Sanchez:21}
giving emphasis on sub-galaxy ($\sim$2\,kpc) scales as dominant in the formation
history of galaxies, with larger scales showing the equivalent of an integrated relation.
Our analysis goes beyond short-timescale relations, such as those probed by emission lines.
The trends with respect to stellar populations depend on substantially longer timescales
and concern the star formation and chemical enrichment histories of galaxies. We also
support the idea that galaxy-wide scales play some role, as suggested by Fig.~\ref{fig:ProbPlot},
but the trends regarding smaller scales should be taken into account in any successful theory
of galaxy formation.

Finally, it did not escape our attention that the correlations with
line strength indicators are stronger than those with physical parameters
such as age or metallicity, derived from population synthesis modelling.
Given that the line strength measurements are observationally produced in an
independent way to velocity dispersion, we propose that the model fitting
process introduces a scatter that does not truly reflect the strong connection
between star formation histories and local velocity dispersion.

\section*{Acknowledgements}
IF acknowledges support from the Spanish Research Agency of the Ministry of Science and Innovation (AEI-MICINN) under grant PID2019-104788GB-I00. MT acknowledges the support of CNPq (process \#307675/2018-1). OL acknowledges STFC Consolidated Grant ST/R000476/1 and a Visiting Fellowship at All Souls College and at the Physics Department, Oxford. RRdC acknowledges the support from FAPESP through grant 2020/16243-3. Funding for SDSS-III has been provided by the Alfred P. Sloan Foundation, the Participating Institutions, the National Science Foundation, and the U.S. Department of Energy Office of Science. The SDSS-III web site is \href{http://www.sdss3.org/}{http://www.sdss3.org/}.

\section*{Data availability}
This work has been fully based on publicly available data: galaxy spectra were retrieved from the SDSS \href{https://www.sdss.org/dr17/}{DR17 archive} and stellar population synthesis models can be obtained from the respective authors.  The excellent MaNGA/Marvin database can be accessed at \href{https://www.sdss4.org/dr17/manga/marvin/}{this link.}

\bibliographystyle{mnras}
\bibliography{MngSpx_subglx}

\appendix

\section{Distribution of line strengths for individual galaxies}
\label{app:single}
\begin{landscape}
  \begin{figure}
    \includegraphics[width=80mm]{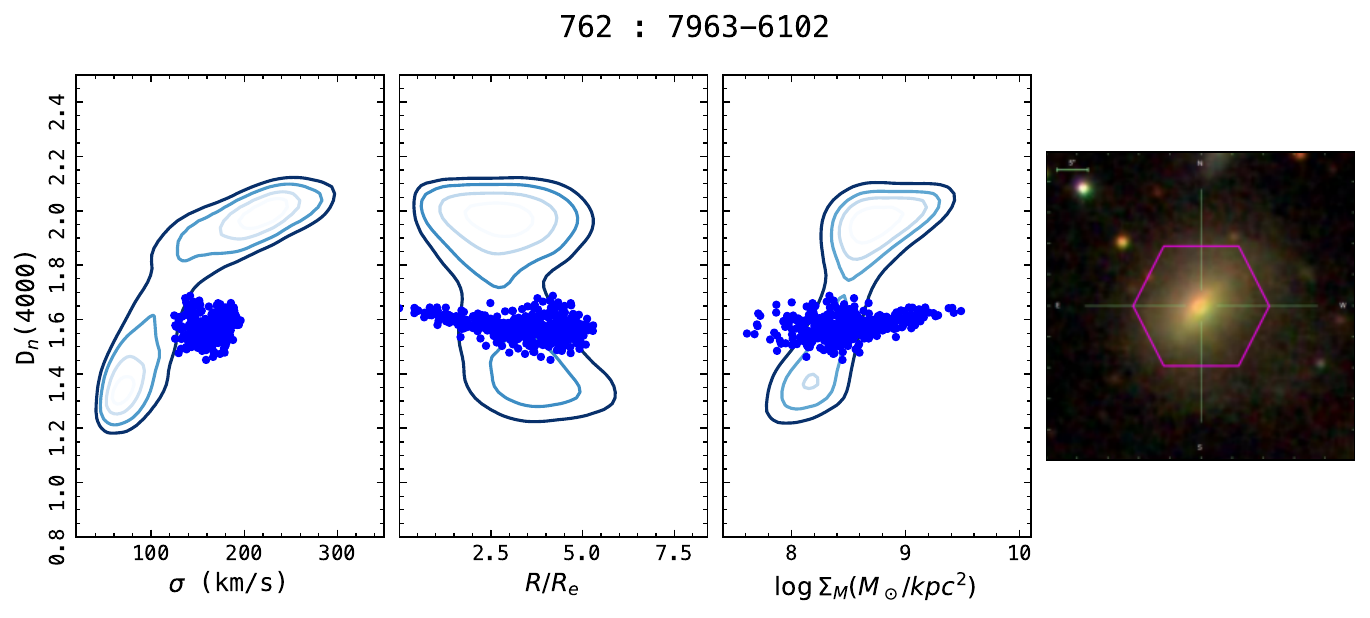} 
    \includegraphics[width=80mm]{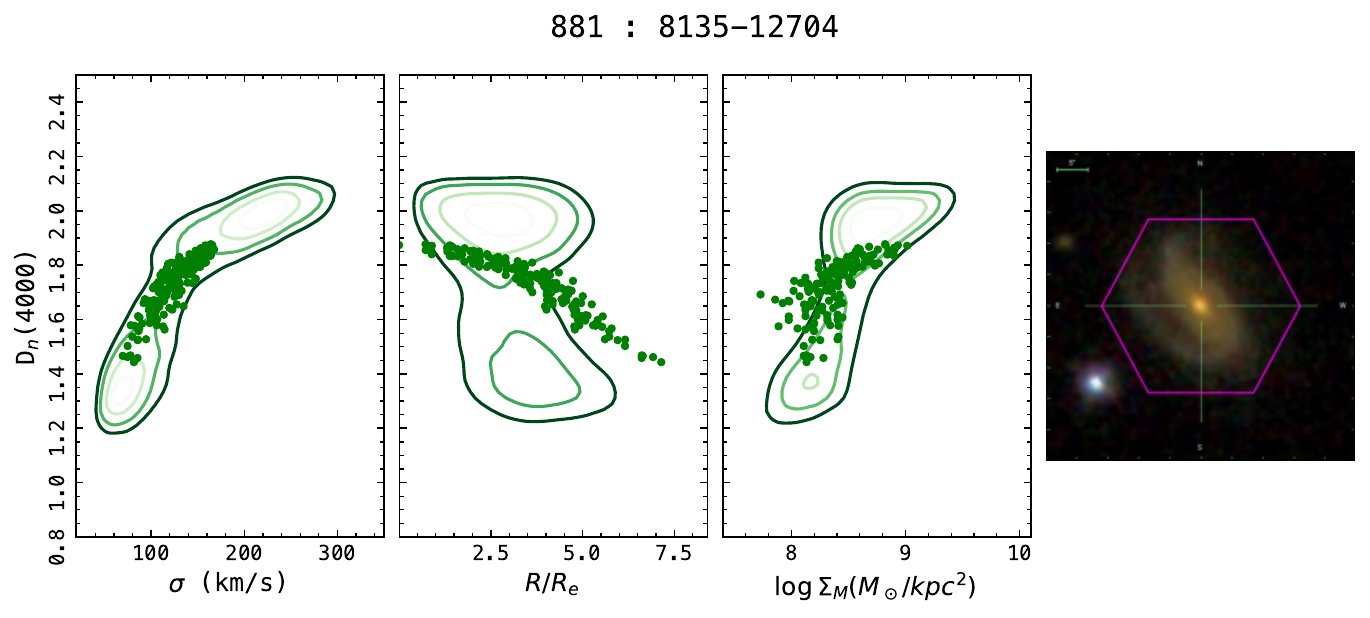} 
    \includegraphics[width=80mm]{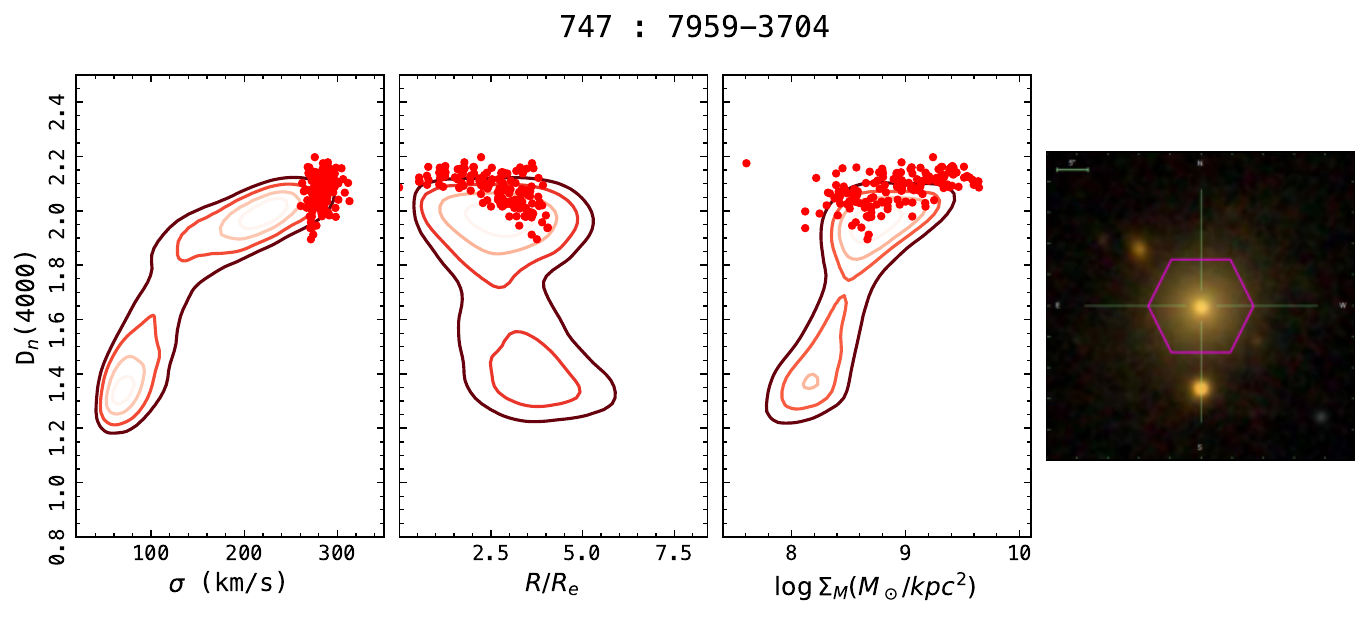} 
    \includegraphics[width=80mm]{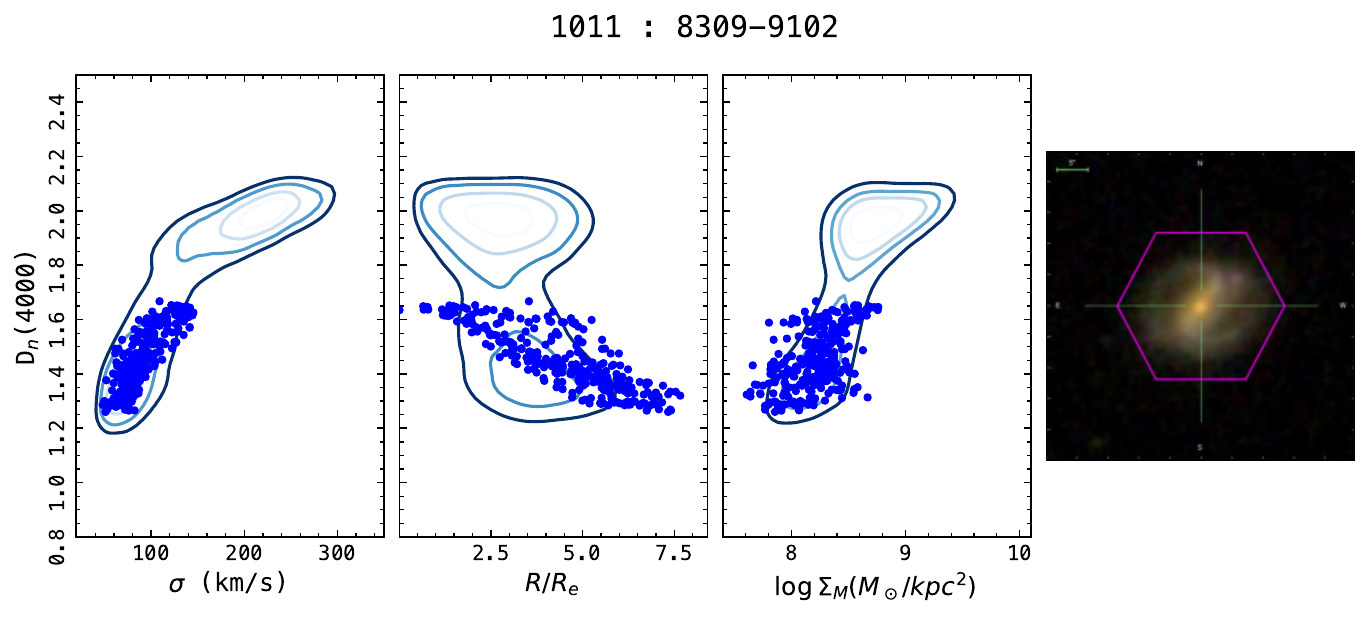} 
    \includegraphics[width=80mm]{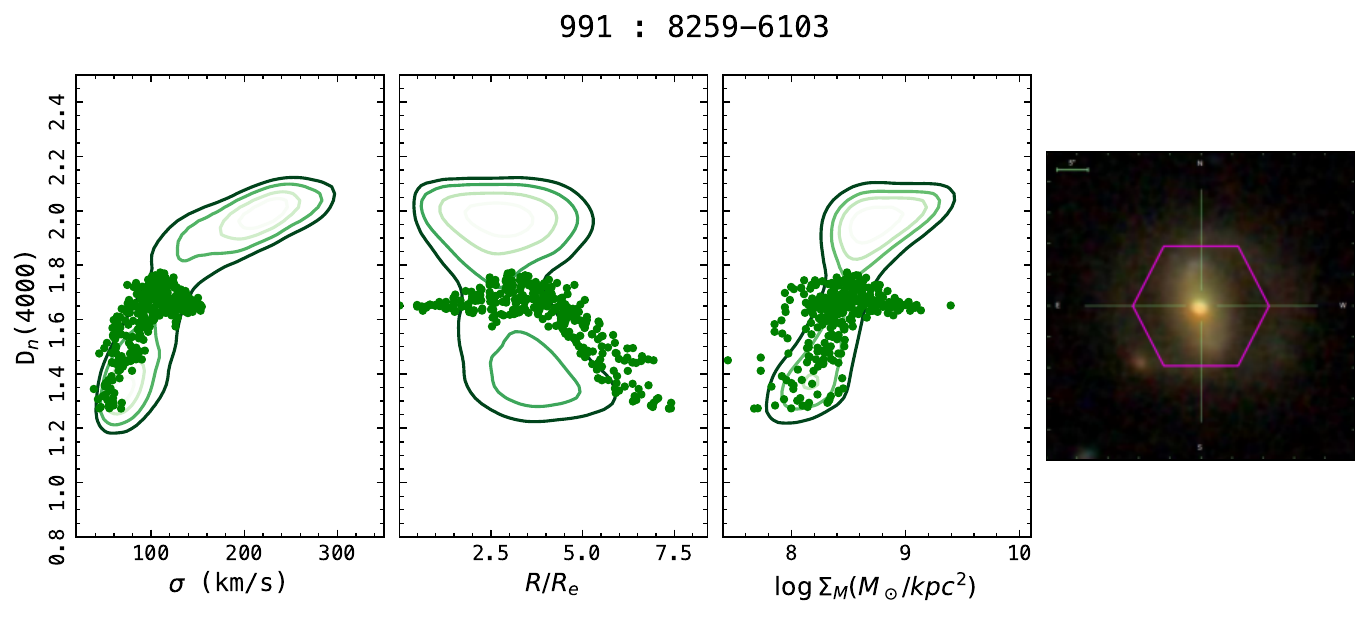} 
    \includegraphics[width=80mm]{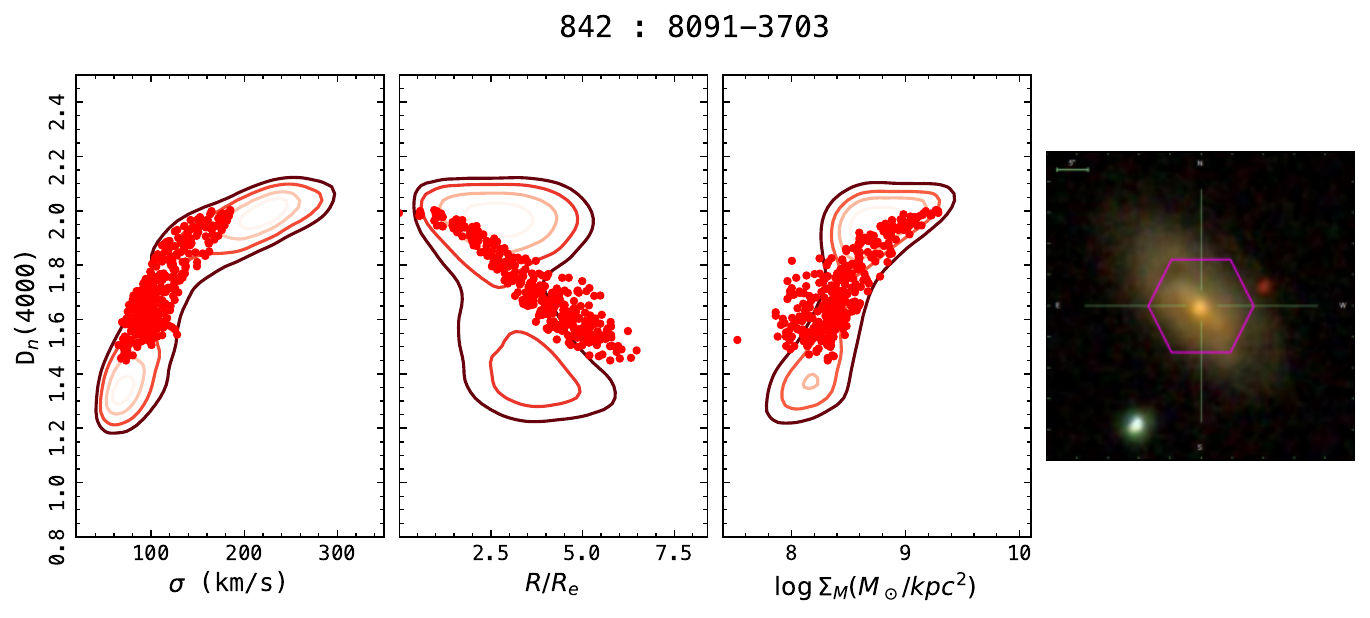} 
    \includegraphics[width=80mm]{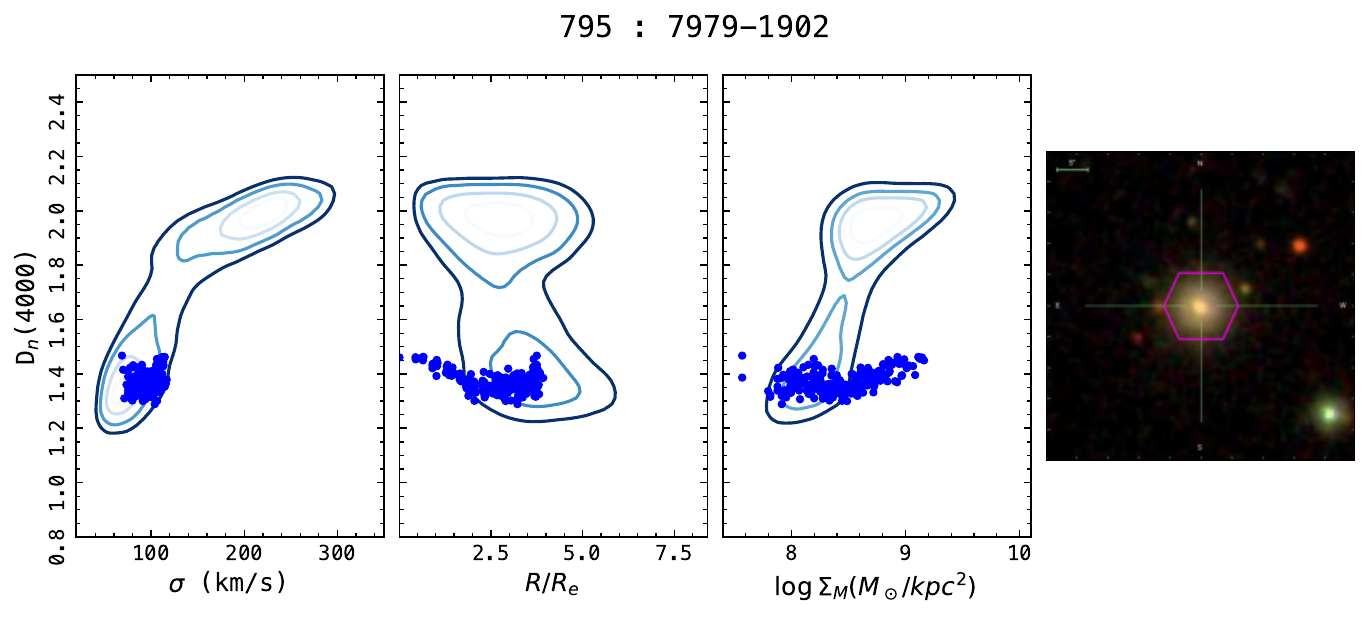} 
    \includegraphics[width=80mm]{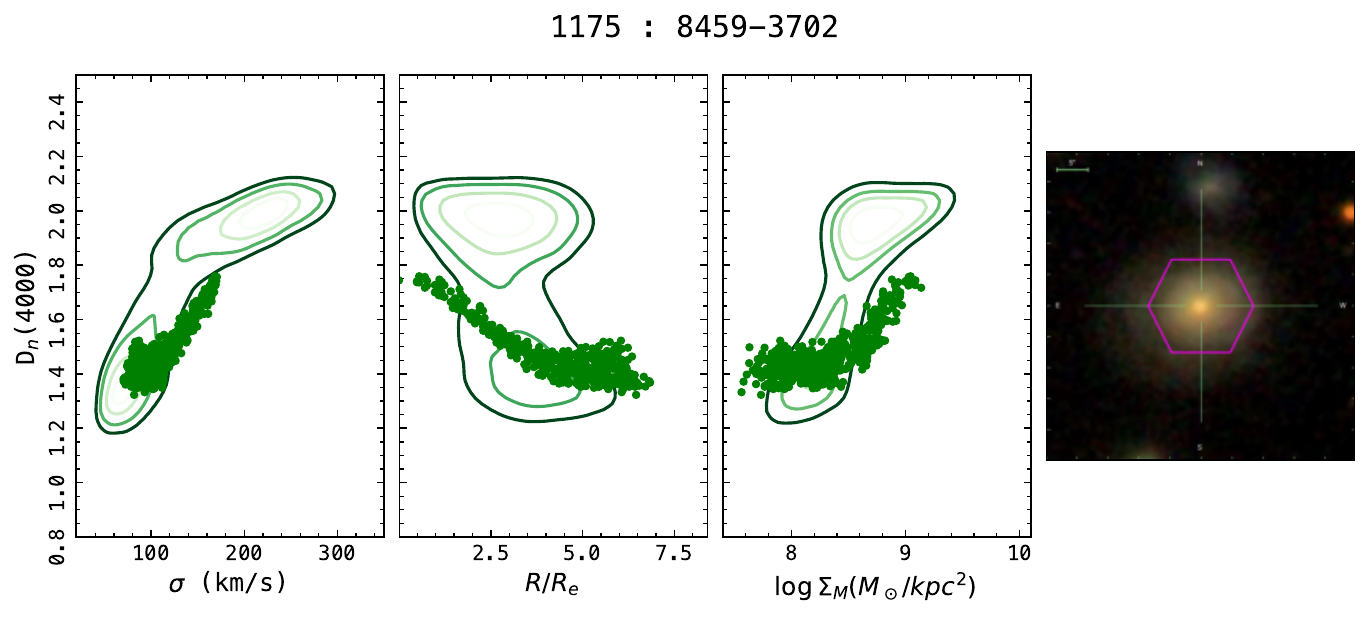} 
    \includegraphics[width=80mm]{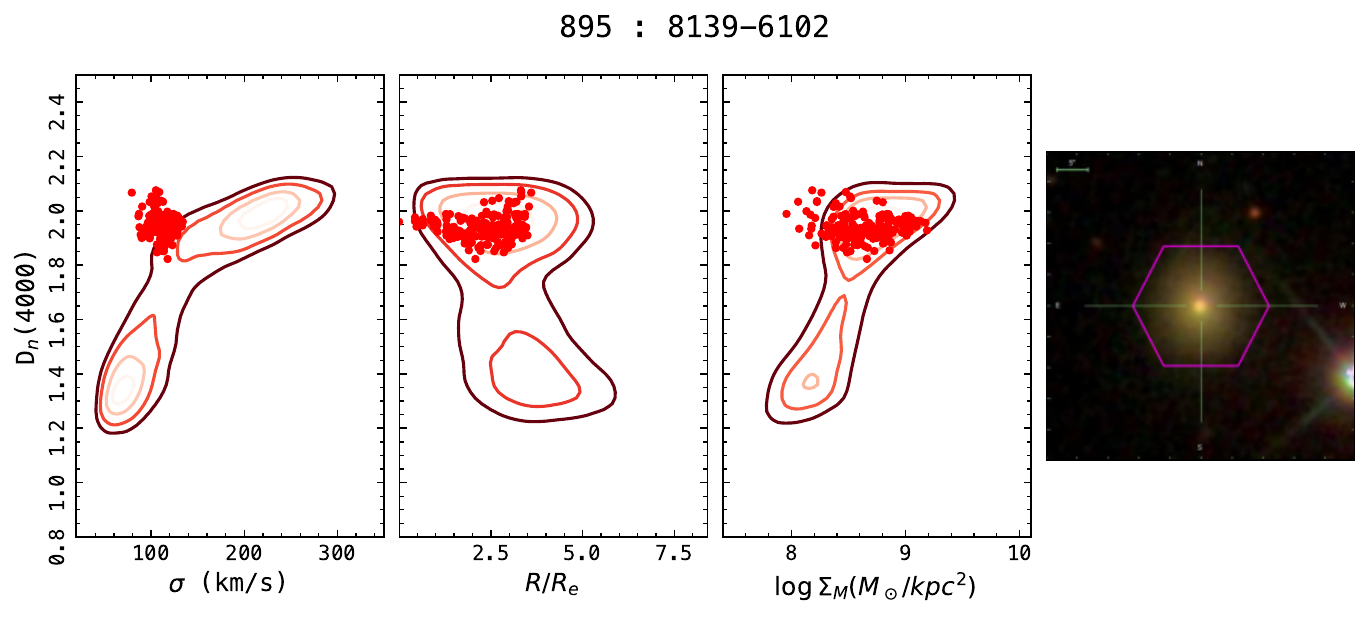} 
    \includegraphics[width=80mm]{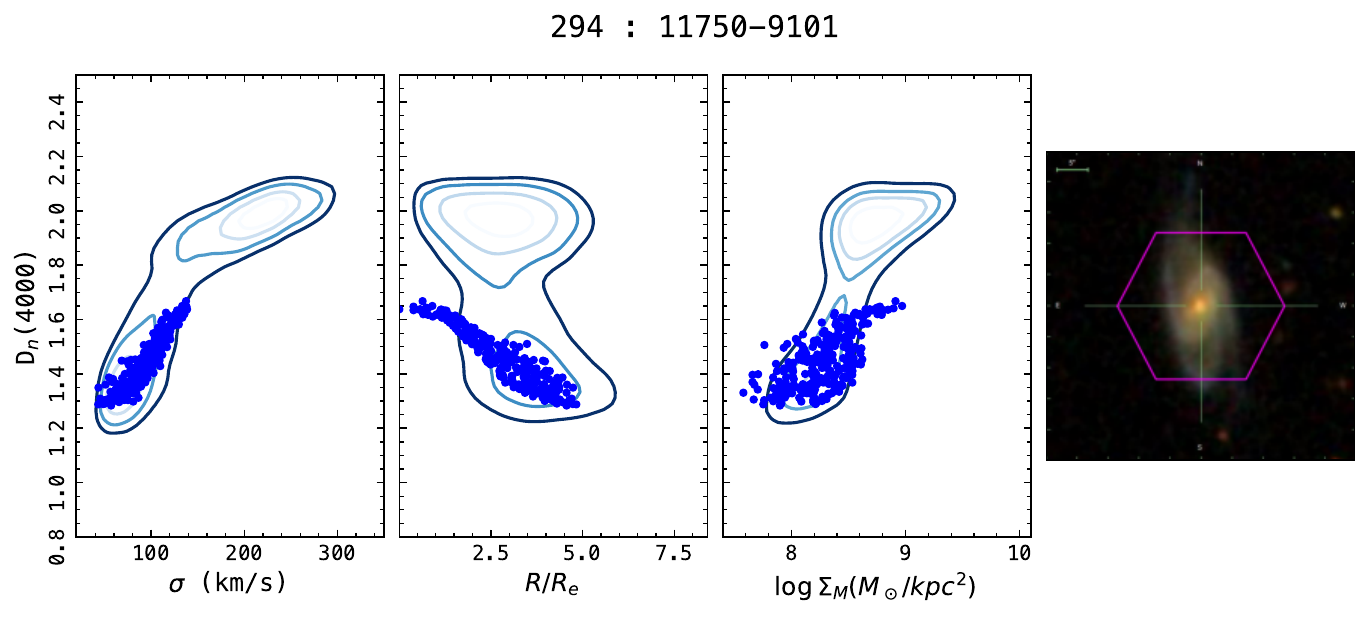} 
    \includegraphics[width=80mm]{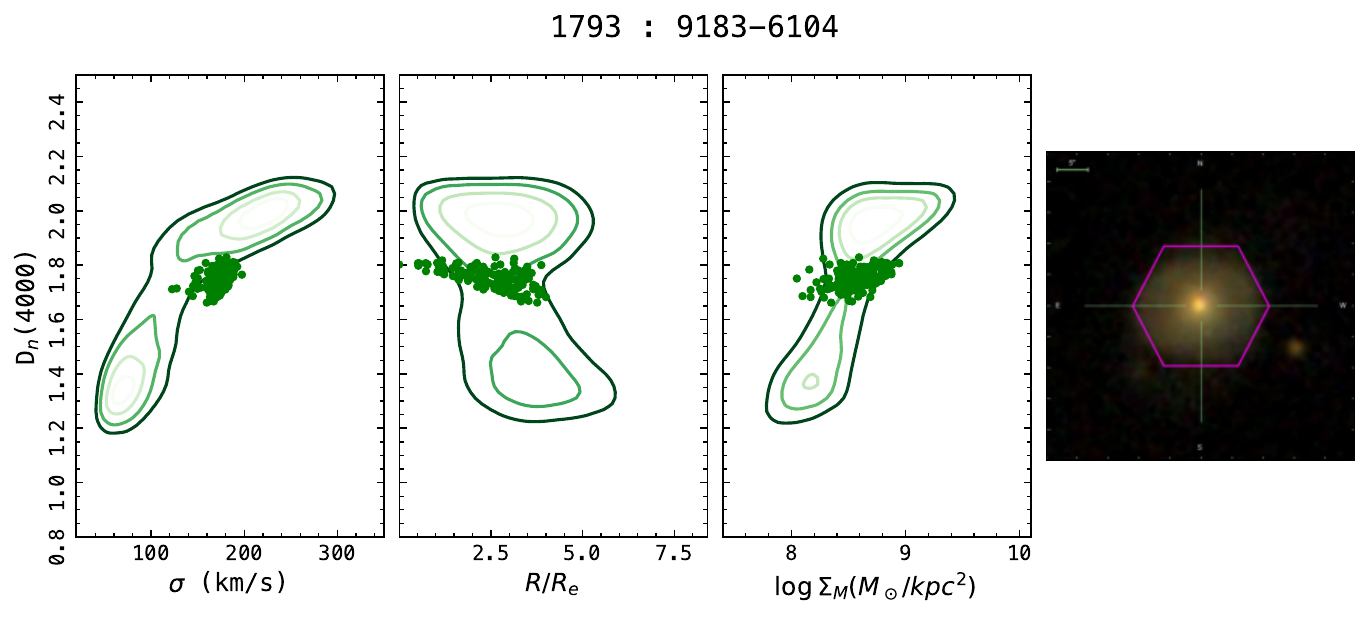} 
    \includegraphics[width=80mm]{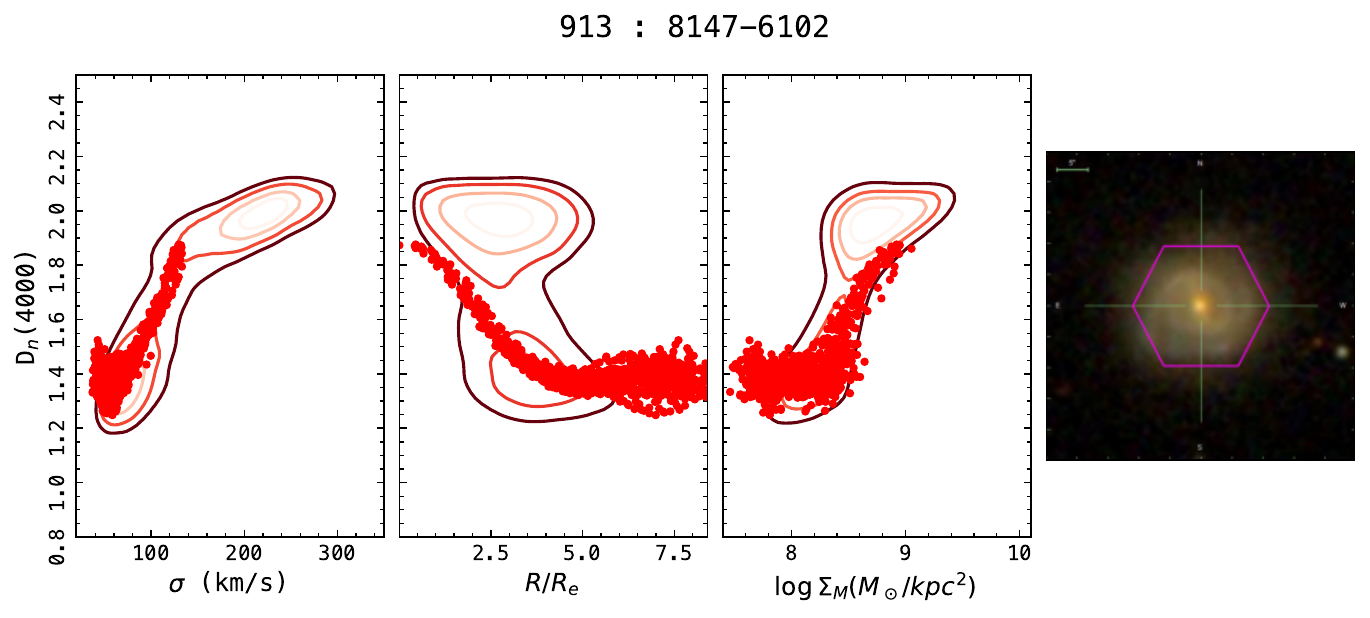} 
    \caption{Distribution of spaxel measurements of $D_n(4000)$ in a small
      selection of galaxies. A stamp of each galaxy is included for reference.
      The subset is colour coded regarding their identification -- from \citet{JA:19} --
    as blue cloud (left), green valley (middle) or red sequence (right).}
    \label{fig:glxs}
  \end{figure}
\end{landscape}

We show in Fig.~\ref{fig:glxs} the distribution of line strengths
measured in a few galaxies from the sample. From left to right they
correspond to representative galaxies in the blue cloud, green valley
and red sequence, respectively. Each diagram includes as coloured dots
the 4000\AA\ break strength with respect to stellar velocity dispersion,
galactocentric distance (measured in fractions of the effective radius),
and stellar mass surface density. Each panel also shows, as contours, the
distribution of the total spaxel sample (i.e. the results of Fig.~\ref{fig:spxALLAbs}). 
For reference, a coloured stamp
is included to the right of each set of panels, including in pink the
footprint of the MaNGA IFU.

\section{Correlation in the outermost spaxels}
\label{app:outer}
\begin{figure*}
\includegraphics[width=160mm]{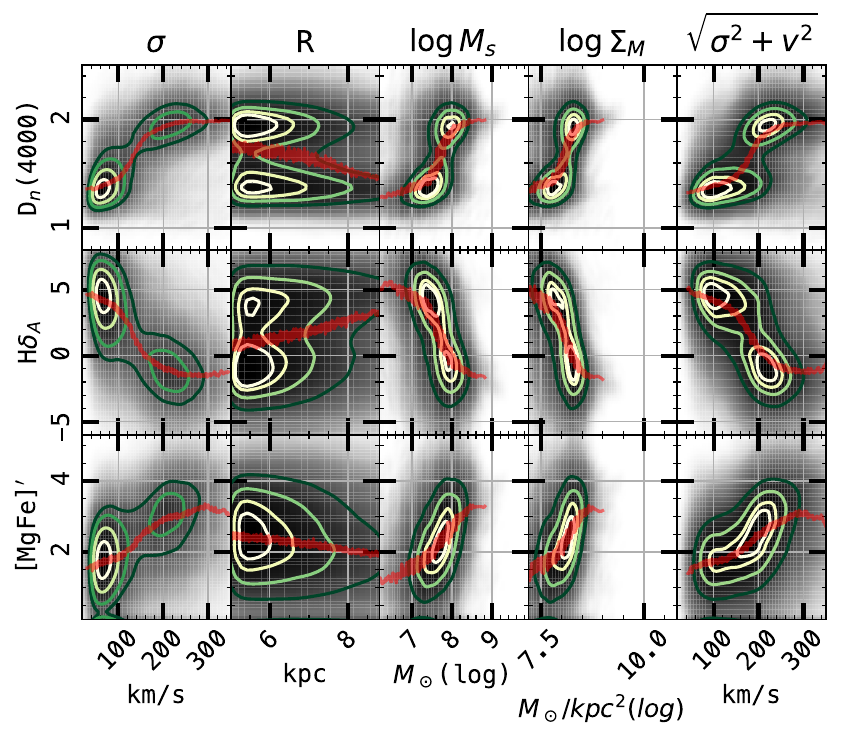}
\caption{Equivalent of Fig.~\ref{fig:spxALLAbs} for the subset of the outermost spaxels, located at galactocentric radial distance R$\geq$5\,kpc.}
\label{fig:spxOuterAbs}
\end{figure*}

One might argue that most of the results in this sample could be mainly caused by the
stellar populations in the central regions, i.e. dominated in most cases by the
bulge component. In fact, the dominant role of velocity dispersion may be reminiscent of
the well-known correlation between supermassive black hole mass and the velocity
dispersion of the bulge \citep{FM:00}. In this appendix we show that the inherent
trend cannot be simply ascribed to the bulge. Fig.~\ref{fig:spxOuterAbs} shows the equivalent of
Fig.~\ref{fig:spxALLAbs} for the subset of spaxels at galactocentric distance R$\geq$5\,kpc
(keeping the S/N theshold above 5). This subset comprises 460,203 spaxels, and the
trends are consistent with the full sample -- of course noting that, by construction,
the range in galactocentric radii is restricted. In addition, we show in Tab.~\ref{tab:corrOuter}
the equivalent of Tab.~\ref{tab:corrAbs} for these outermost spaxels. We conclude that
neither the limited spatial resolution (PSF) nor the role of the bulge is significant
in the trends found in this paper. Incidentally, note that the bimodality in the
population indicators appears quite strongly at all of the outer radii, whereas the line
strengths depend sensitively on the velocity dispersion.

\begin{table*}
  \centering
  \caption{Correlation coefficients between stellar population line strengths and observables measured in an absolute way in spaxels with S/N$\geq$5 in the outermost spaxels (R$\geq$5\,kpc, see fig~\ref{fig:spxOuterAbs})}
  \begin{tabular}{cccccc} 
    \hline
    Index & $\sigma$ & $R$ & $\log M_s$ & $\log\Sigma_M$ & $\sqrt{\sigma^2+v^2}$\\
    \hline
    \multicolumn{6}{c}{Correlation coefficient}\\
    \hline
    D$_n$(4000)    & $+$0.766$\pm$0.001 & $-$0.193$\pm$0.001 & $+$0.554$\pm$0.001 & $+$0.534$\pm$0.001 & $+$0.569$\pm$0.001\\
    H$\delta_A$    & $-$0.672$\pm$0.001 & $+$0.193$\pm$0.001 & $-$0.501$\pm$0.001 & $-$0.510$\pm$0.001 & $-$0.496$\pm$0.001\\
    $[{\rm MgFe}]^\prime$ & $+$0.513$\pm$0.001 & $-$0.205$\pm$0.001 & $+$0.425$\pm$0.001 & $+$0.421$\pm$0.001 & $+$0.334$\pm$0.002\\
    \hline
    \multicolumn{6}{c}{Standard deviation (and 1$\sigma$ error)}\\
    \hline
    D$_n$(4000)    & 0.175$\pm$0.031 & 0.289$\pm$0.013 & 0.240$\pm$0.030 & 0.242$\pm$0.032 & 0.224$\pm$0.043\\
    H$\delta_A$    & 2.279$\pm$0.235 & 3.141$\pm$0.102 & 2.728$\pm$0.322 & 2.688$\pm$0.330 & 2.654$\pm$0.345\\
    $[{\rm MgFe}]^\prime$ & 0.911$\pm$0.057 & 1.046$\pm$0.054 & 0.962$\pm$0.058 & 0.960$\pm$0.068 & 0.969$\pm$0.069\\
    \hline
  \end{tabular}
  \label{tab:corrOuter}
\end{table*}

\bsp	
\label{lastpage}

\end{document}